\documentstyle[12pt]{article}
\begin{document}
\begin{center}{\Large {\bf Dynamic-Symmetry-Breaking {\it Breathing}
 and {\it Spreading}
Transitions in Ferromagnetic Film Irradiated by Spherical Electromagnetic Wave}}\end{center}

\vskip 1cm

\begin{center}{\it Muktish Acharyya}\\
{\it Department of Physics, Presidency University}\\
{\it 86/1 College Street, Calcutta-700073, India.}\\
{E-mail:muktish.physics@presiuniv.ac.in}\end{center}

\vskip 1cm

\noindent The dynamical responses of a  
ferromagnetic film to a propagating 
spherical electromagnetic wave passing through it are studied by Monte
Carlo simulation of two dimensional Ising ferromagnet. 
For a fixed set of values of the 
frequency and wavelength of the spherical EM wave, and depending on the
values of amplitude of the EM wave and temperature of the system, three
different modes are identified. The static {\it pinned} mode, 
the localised dynamical
{\it breathing} mode and extended dynamical {\it spreading} mode are 
observed. The nonequilibrium 
dynamical-symmetry-breaking {\it breathing} and {\it spreading}
phase transitions are also observed and the transition temperatures are
obtained as functions of the amplitude of the magnetic field of EM wave.
A comprehensive phase diagram is drawn. The boundaries of breathing and
spreading transitions merge eventually at the 
equilibrium transition temperature for
two dimensional Ising ferromagnet as the value of the amplitude of the 
magnetic field becomes vanishingly small. 

\vskip 3cm


\newpage

\noindent {\bf I. Introduction}

Ising model is a widely used prototype to study the phase transitions
phenomena. Because of its simplicity, even in the case of nonequilibrium 
behaviour, this model is being
extensively and successfully used\cite{reviews}. 
The hysteresis and dynamical phase
transitions are two important responses of kinetic Ising ferromagnet
to an oscillating magnetic field
and play an important role in modern research of nonequilibrium phenomena\cite{reviews}.

Particularly, in this field, the journey was started to study the responses
of kinetic Ising model to an oscillating magnetic field. The dynamical
meanfield equation was solved and dynamic phase transition was observed
\cite{tom}.
Much effort was devoted to study the responses by Monte Carlo simulations.
The hysteretic responses and dynamic symmetry breaking nonequilibrium phase
transition were studied extensively. A considerable amount of research was
performed to establish this phase transition as a nonequilibrium phase
transition\cite{ma1,ma2,ma3,ma4,ma5,rik1,rik2,rik3,rik4}. Recently, the surface and bulk
dynamic transitions were studied in kinetic Ising model and the different
classes of universality were observed\cite{park}.

Not only in Ising ferromagnet, the dynamic phase transition was observed
in Ising metamagnet both from meanfield study\cite{other-meta} and 
Monte Carlo simulations\cite{ma-meta}.
The various kinds of nonequilibrium phase transitions were observed in
classical vector spin models. Recently, the nonequilibrium phase 
transition was observed\cite{vatansever} in magnetic nanocomposites by MC simulations.

Apart from the simulational studies of kinetic Ising model, researchers are also
paying attention to observe the dynamical phase transitions in Blume-Capel model\cite{blume},
Blume-Emery-griffith model\cite{beg} and
classical vector spin models\cite{ijmpc,jung,hall,mdt}. 

Experimentally, the dynamic symmetry breaking was also observed in ultrathin
Co film on Cu(001) by surface magneto-optic Kerr effect\cite{expt}. However, it may be
mentioned here that all the studies referred above have a common feature.
In those cases, the magnetic field was sinusoidally oscillating but was
uniform over the space at any particular instant. 

Very recently, the nonequilibrium dynamic phase transition was also observed
in Ising ferromagnet swept by linearly polarised electromagnetic 
{\it plane} wave\cite{prop,rfim,polarised}. 
In these cases, the
spatio-temporal variations of the magnetic field were considered. Here, the
coherent motion of spin clusters was found and phase boundaries were drawn.

In this present article, the dynamical responses of two dimensional Ising
ferromagnet to a electromagnetic {\it spherical} wave are studied by Monte Carlo
simulation. 
The layout of the paper is as follows: section-II decribes
the model and the Monte Carlo simulation scheme, the numerical results
are reported in section-III, the paper ends with concluding remarks in 
section-IV.

\vskip 1cm

\noindent {\bf II. Model and simulation}

The two dimensional Ising ferromagnet (having uniform nearest neighbour
interaction)
in presence of a 
propagating spherical electromagnetic field wave 
(having spatio-temporal variation) can be represented by the following time
dependent Hamiltonian,
\begin{equation}
H(t) = -J\Sigma s(x,y,t) s(x',y',t) 
-\Sigma h(x,y,t) s(x,y,t)
\end{equation}
\noindent 
The Ising spin variable, $s(x,y,t)$ assumes value $\pm 1$ at lattice
site $(x,y)$ at time $t$ on a square lattice of linear size $L$.
The uniform ferromagnetic nearest neighbour interaction strength is $J(>0)$. 
The first sum
represents the Ising spin-spin interaction. The spin-field interaction 
resides in the second summation.
The $h(x,y,t)$ is the value of the magnetic field (at point $(x,y)$ and at
any time $t$) of the propagating (radially outward)
 spherical electromagnetic wave (originating from the centre($x_0,y_0$) of the lattice). 
The form of spherically propagating wave is
\begin{equation}
h(x,y,t) = h_0 {{{\rm e}^{i(2\pi f_0 t -2\pi r/\lambda)}} \over r}
\end{equation}
Where, $r=\sqrt{(x-x_0)^2 + (y-y_0)^2}$.
The $h_0$, $f_0 = {{\omega_0} \over {2\pi}}$ 
and $\lambda$ represent the amplitude, frequency and the
wavelength respectively of the propagating spherical electromagnetic field wave which
originates from $r=0$ and propagates radially outwards. This form of the propagating field
was obtained from the solution of spherically symmetric Maxwell's equation
representing the electromagnetic wave.
In the present simulation, a $L\times L$ square lattice is considered. 
The boundary condition, used here, is periodic
in both ($x$ and $y$) directions. 
The initial ($t=0$) configuration, as the all
spins are up 
($s(x,y,t=0) = +1$ for all $x$ and $y$)), is taken here. 
The spins are updated
randomly (a site ($x,y$) is chosen at random) and spin flip occurs 
(at temperature $T$)
according to the Metropolis probability\cite{binder} of single spin flip ($W$)
\begin{equation}
W(s \to -s)={\rm Min}[{\rm exp}(-{\Delta E}/{kT}),1],
\end{equation}
where $\Delta E$ is the change in energy due to spin flip and $k$ is
Boltzmann constant. $L^2$ such random updates of spins defines the unit
time step here and is called Monte Carlo Step per spin (MCSS).
Here, the magnetic field and the temperature are measured in the units of $J$
and $J/k$ respectively.
The dynamical steady state is reached by heating the system slowly (in presence of
the propagating field)
in small step ($\delta T = 0.05$ here) of temperature.
It may be mentioned here that the same dynamical steady state was observed to be
achieved by cooling
the system from a high temperature random configuration.
The
frequency and wavelength of the propagating magnetic field were kept fixed 
($f=0.01$ and $\lambda=15.0$) throught the
study. The total length of simulation is $2\times {10^5}$ MCS 
and first $10^5$ MCS
transient data were discarded to achieve the stable dynamical steady state. 
Since the frequency of the propagating field is $f=0.01$, the complete cycle
of the field requires 100 MCS. So, in $10^5$ MCS, $10^3$ numbers of cycles of
the propagating field are present. The time averaged data over the full cycle
(100 MCSS) of the propagating field are further averaged over 1000 cycles.
Here, the number of cycles is denoted by $n_c$.

The quantities measured are:
Instantaneous local magnetisation density in the 
circle of radius $\lambda/2$ is $m_b(t)=\sum s(x,y,t)/(N_b)$, where the sum is
carried over the number of sites ($N_b$)
 lying within the circle of radius $\lambda/2$
centered at the centre ($x_0,y_0$) of lattice. In the present study, the lattice size $L$ is 
taken equal to 101. So, the coordinates of the centre is ($x_0$=51,$y_0$=51).
and $N_b$ is the total number of lattice sites within this circle. 
The dynamic
order parameter of breathing transition 
is defined as $Q_b = {{\omega} \over {2\pi}} \oint
m_b(t) dt$. 
The fluctuations in dynamic order (for breathing) $<{\delta Q_b}^2>=
<{Q_b}^2> - <{Q_b}>^2$. The average value of $Q_b$ and the fluctuation are
 calculated over $n_c=1000$ number of random samples. It is also checked that
this number of samples is sufficient to have the data within the specified
accuracy.
The Fourier component ${\hat m_b(\omega)} = {{\omega_0}
 \over {2\pi n_c}}\int_{0}^{{2\pi n_c} \over {\omega_0}} m_b(t) 
e^{i\omega t} dt$. 
The integrated power for breathing $P_b = \int_{{\omega_0} \over 2}^{{3\omega_0} \over 2} 
{\hat m_b(\omega)} d\omega$. In general, ${\hat m_b(\omega)}$ and $P_b$ are
complex.

The quantities to study the spreading transition are:
The boundary touching ring magnetisation density
 $m_s(t) = \sum s(x,y,t)/N_s$, where
the summation is carried over the sites residing within the ring of width
$\lambda/2$ touching the boundary, and $N_s$ is the total number of lattice
sites within that ring. 
It is a region bounded between two concentric ($x_0=51,y_0=51$) 
circles having outer radius $r_{out}=L/2$ and
the inner radius $r_{in}=L/2-\lambda/2$.
The dynamic
order parameter of spreading transition $Q_s = {{\omega} \over {2\pi}} \oint
m_s(t) dt$. 
The fluctuation in dynamic order (for spreading) $<{\delta Q_s}^2>=
<{Q_s}^2> - <{Q_s}>^2$.
The Fourier component ${\hat m_s(\omega)} = {{\omega_0} \over {2\pi n_c}}\int_{0}^{{2\pi n_c} \over {\omega_0}} m_s(t) 
e^{i\omega t} dt$. 
The integrated power for spreading $P_s = \int_{{\omega_0} \over 2}^{{3\omega_0} \over 2} 
{\hat m_s(\omega)} d\omega$. Here also, ${\hat m_s(\omega)}$ and $P_s$ are 
complex, in general.

\vskip 1cm

\noindent {\bf III. Simulational results:}

A square lattice of size $L=101$ is considered. At the centre 
($x_0=51,y_0=51$) there is a source of spherical electromagnetic wave 
propagating radially outward. The initial configuration of the system was taken where
all the Ising spins, $s(x,y,t=0)$, assume the value $+1$. The system is heated
(slowly in the step $\Delta T = 0.05$) in presence of this
field. For low enough temperature, i.e. $T=0.3$ (and 
$h_0=2.5$, $f_0=0.01$, $\lambda
=15.0$) it is observed that no spin flip occurs and a dynamically pinned
phase is observed (Fig-1(a)). On the other hand, at the temperature 
$T=1.45$,(keeping all the other parameter fixed), 
considerable number of flipping of the spins starts to occur
 near the centre, since
the values of the magnetic fields are higher near the centre. 
Fig-1(b), shows  
snapshot of a typical breathing mode, where the local magnetisation
$m_b(t)$ starts to oscillate. 

The dynamics of a typical breathing mode is pictorially represented by 
Fig-2. Here, the snapshots at two different times are shown
for a fixed set of values of $T=1.25$, $h_0=2.5$, $f=0.01$ and
$\lambda=15.0$. Fig-2(a), shows the spin configuration at instant $t=3970$
MCSS and Fig-2(b) shows the same in a later instant $t=4000$ MCSS. From
this figure, the dynamical breathing mode, is evident. A central and 
{\it localised} dynamical mode observed here, is called the {\it breathing} mode.

To distinguish the dynamical {\it breathing} mode from the dynamical
{\it spreading} mode, a typical high temperature and high field dynamical
mode is shown in Fig-3. For $T=2.30$ and $h_0=10.0$, the spin wave propagation
is shown and it represents a propagation of spherical spin waves which 
{\it spreads} through the entire lattice. This is certainly a 
{\it nonlocalised or spreading} dynamical mode remarkably distinct from the
{\it localised or breathing} mode.

As a result, in the {\it breathing} mode the instantaneous local magnetisation
density $m_b(t)$ becomes a periodic function of time 
but oscillates asymmetrically
about a nonzero avarage value. One such plot of $m_b(t)$ versus $t$ is shown
in Fig-4(a) for the values of the parameter $T=0.5$, $h_0=2.5$. Keeping the
value of $h_0=2.5$ fixed, the symmetric oscillation of $m_b(t)$ is observed
at higher temperature ($T=1.25$)
and shown in Fig-4(b).  
One gets a transition, called the {\it breathing} transition, from low
temperature dynamically symmetry broken phase 
to a high temperature dynamically symmetric phase.

Now, let us see what happens in the case of spin wave {\it spreading}
 mode. Whether
the spin wave spreads over the entire lattice or not, it must be confirmed
by studying the time variation of the magnetisation density within the ring
(of finite width) touching the boundary of the lattice. This promted to define
the ring magnetisation density in the way described in the last section
(section-II). Here also, the boundary touching ring magnetisation density
$m_s(t)$ is plotted against the time $t$ and shown in Fig-5. For $h_0=2.5$
and $T=2.0$ the symmetry broken oscillation of $m_s(t)$ is observed and
shown in Fig-5(a). Increasing the temperature to $T=2.5$ (keeping $h_0=2.5$ 
fixed), the oscillation of $m_s(t)$ becomes symmetric (about $m_s(t)=0$ line).
Here also, one gets a transition, called the {\it spreading} 
transition, from low
temperature dynamically symmetry broken phase 
to a high temperature dynamically symmetric phase.

From the plots of $m_b(t)$ (in Fig-4) and $m_s(t)$ (in Fig-5), the high 
temperature  symmetric and low temperature symmetry broken modes of oscillation
were observed. These oscillations are {\it periodic} in nature. 
This periodicity is confirmed by studying the Fourier 
components ${\hat m_b(\omega)}$ and ${\hat m_s(\omega)}$ as 
the functions of $\omega$. These are called
the power spectra of $m_b(t)$ and $m_s(t)$. The power spectra for the breathing
transition is shown in Fig-6 and that for {\it spreading} transition is shown
in Fig-7. From the figures, it is clear that the spin wave {\it breathing} and
{\it spreading} modes are periodic (with the same periodicity,i.e., $f=0.01$ of the spherical
electromagnetic wave) in both (symmetric and symmetry-broken) phases.

To study the symmetry broken dynamic phase transitions (both breathing and
spreading) quantitatively and to estimate the transition temperatures
precisely, one has to define the order-parameters for the transitions.
The time averaged magnetisation density over a full cycle of the EM field 
defines the dynamic order parameter. By definition, the symmetry broken phase
will yield nonzero value of the dynamic order parameter and hence represent the
dynamically ordered phase. The disordered phase (vanishingly small value of
the order parameter) is dynamically symmetric phase. The temperature dependence
of the dynamic order parameters $Q_b$ (for breathing) and $Q_s$ (for spreading)
are plotted in Fig-8 for two different values of $h_0$ 
(equals to 2.5 in Fig-8(a) and
equals to 3.5 in Fig-8(b)).

To have a precise measure of the dynamic {\it breathing} and {\it spreading} transitions
the variances $<(\delta Q_b)^2>$  and $<(\delta Q_s)^2>$ of $Q_b$ and $Q_s$ 
respectively are studied as the function of temperature and shown in Fig-9. 
Here the peaks of $<(\delta Q_b)^2>$ and $<(\delta Q_s)^2>$
indicate the respective transitions. 
In Fig-9(a),
the plots are shown for $h_0=2.5$. 
The {\it breathing} and {\it spreading}
transitions occur at $T_b=0.95$ and $T_s=2.25$ respectively. 
Similar plots 
for $h_0=3.5$ are shown in Fig-9(b). 
In this case, the {\it breathing} and {\it spreading}
transitions occur at $T_b=0.75$ and $T_s=2.20$ respectively. 
It may be mentioned at here that the small step of temperature, by which the
system was heated from a perfectly ordered configuration, is $\Delta T =0.05$.
So, the transition temperature has the maximum error of amount $\pm 0.05$.

The absolute value of integrated powers, (obtained from power spectra) for
breathing and spreading modes, are studied also as a function of temperature
and shown in Fig-10. The integrated powers $|P_b|$ for breathing mode and
$|P_s|$ for spreading mode are plotted against the temperature $T$ in Fig-10(a)
for $h_0=2.5$ and in Fig-10(b) for $h_0=3.5$ respectively. From the figures it
is clear that the breathing transition is also indicated approximately by the sharp peak
of $|P_b|$. However, the integrated power $|P_s|$ for the spreading transition
shows a relatively smeared peak above the spreading transition temperature $T_s$.

Obtaining the breathing and spreading transition temperatures for different values
of the amplitude ($h_0$) of the spherical EM field, a comprehensive phase diagram
is drawn in the plane formed by $h_0$ and $T$. This phase diagram is shown in
Fig-11. Three distinct phases, the {\it pinned}, the {\it breathing} 
and the {\it spreading} and their boundaries are shown
(with maximum errors in measuring the transition temperatures). The boundaries representing
the breathing
and spreading transitions come closer as $h_0$ decreases and eventually reach the
limit $T=2.269...$ (the Onsager's value) as $h_0 \to 0$. 

\newpage

\noindent {\bf IV. Concluding remarks:}

The dynamical responses of two dimensional Ising ferromagnet 
irradiated by a spherical electromagnetic wave are
studied by Monte Carlo simulation. For very low temperature and small values of the amplitude
of EM wave, the system remains in a pinned state. As the system is heated the spins in the
central regions starts to flip and a very localised dynamical breathing mode was observed. 
The local magnetisation density starts to oscillates. A dynamical phase transition, called
breathing transition associated with a symmetry breaking is observed. For high value of the
temperature of the system the collective spin wave (spherically symmetric) starts to propagate
and spreads over the entire lattice. A dynamical spreading mode was observed. Here also a 
dynamic symmetry breaking spreading transition is observed. These breathing and spreading
transition temperatures depend on the amplitude of the EM wave. A comprehensive phase 
diagram is drawn in the plane formed by the temperature and the amplitude of the EM wave.
The boundaries of breathing and spreading transitions come closer as the amplitude of the
EM wave decreases and eventually merge at the equilibrium ferro-para transition point for
the two dimensional Ising ferromagnet as the value of the amplitude approaches zero.

The dynamic symmetry breaking {\it breathing} and {\it spreading} transitions are new
kinds of nonequilibrium phase transitions observed in two dimensional ising ferromagnet
irradiated by spherical electromagnetic wave. There are further scopes of studies in
this front. One may study it in three dimensions also. The detail study of this phase
transition and scaling analysis would lead to have an idea about the universality class
of these transitions. It would be interesting and supportive to study
these transitions experimentally using ultrathin Co film irradiated by strong EM wave
by surface magneto optic Kerr effect.

\vskip 1cm

\noindent {\bf Acknowledgements:} Author would like to thank the library facilities provided
by the University of Calcutta.

\newpage 
\setlength{\unitlength}{0.240900pt}
\ifx\plotpoint\undefined\newsavebox{\plotpoint}\fi
\sbox{\plotpoint}{\rule[-0.200pt]{0.400pt}{0.400pt}}%


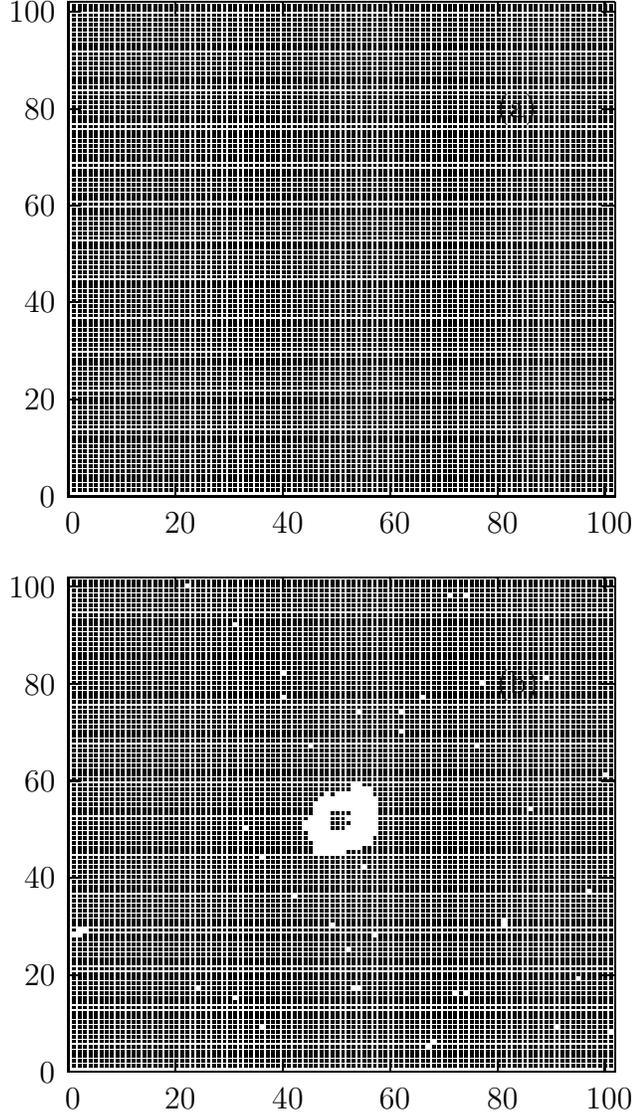
\noindent {\bf Fig-1.} The pinned (a) and breathing (b) modes
on the lattice. Dots represents up spins only. (a) $h_0=2.5$, $T=0.30$ and
(b) $h_0=2.5$ and $T=1.45$. Here, $\lambda=15.0$, $f_0={{\omega_0} \over {2\pi}}=0.01$ and $t=4000$ MCSS.
\newpage
\setlength{\unitlength}{0.240900pt}
\ifx\plotpoint\undefined\newsavebox{\plotpoint}\fi
\sbox{\plotpoint}{\rule[-0.200pt]{0.400pt}{0.400pt}}%
%

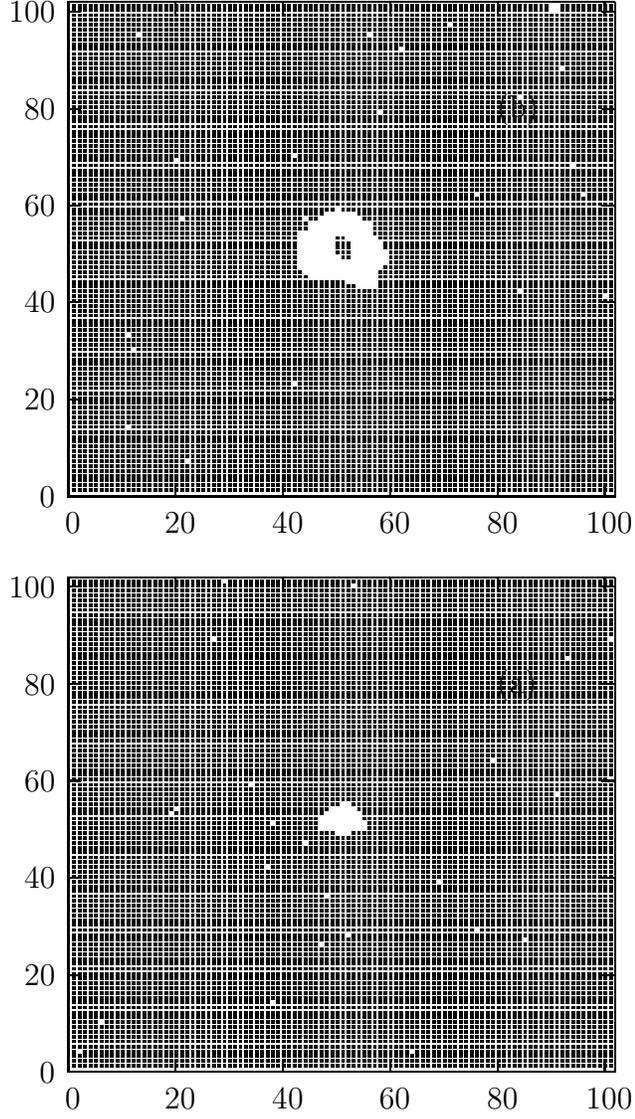
\noindent {\bf Fig-2.} The breathing mode on the lattice. 
(a) $h_0=2.5$, $f_0=0.01$, $\lambda=15.0$, $T=1.25$ at $t=3970$ MCSS.
(b) $h_0=2.5$, $f_0=0.01$, $\lambda=15.0$, $T=1.25$ at $t=4000$ MCSS.

\newpage
\setlength{\unitlength}{0.240900pt}
\ifx\plotpoint\undefined\newsavebox{\plotpoint}\fi
\sbox{\plotpoint}{\rule[-0.200pt]{0.400pt}{0.400pt}}%
%

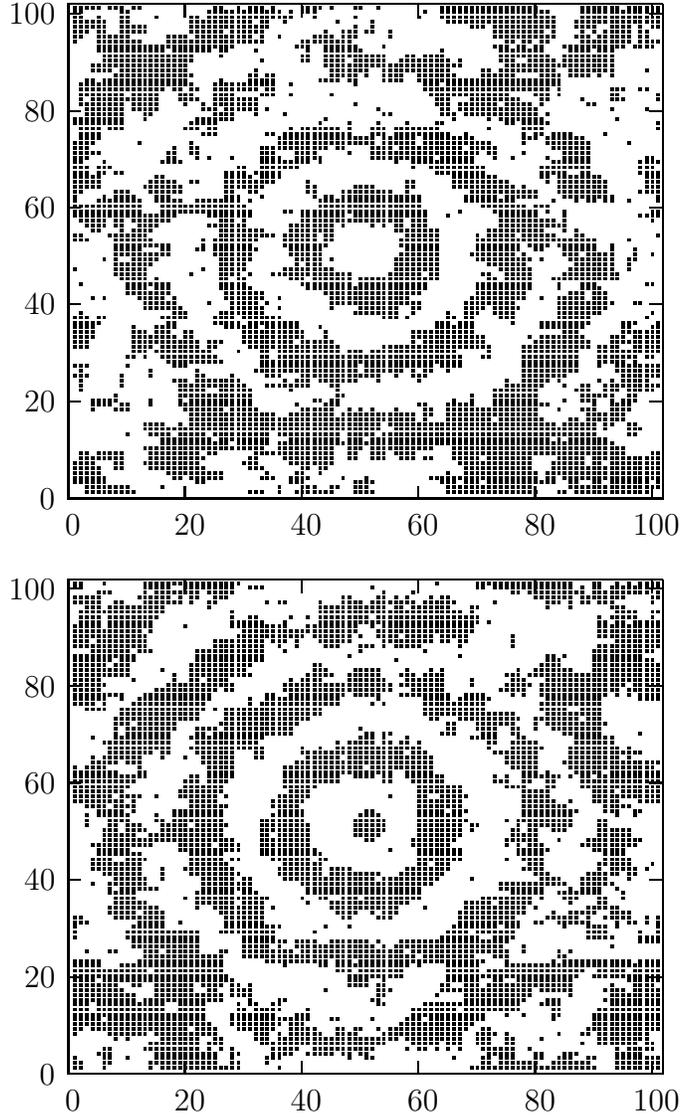
\noindent {\bf Fig-3.} A high field ($h_0 = 10.0$)-high temperature ($T=2.30$) spin wave propagation mode. 
The up spins ($S_i = +1$) are shown by black dot. The bottom
one shows a spin configuration at 3970 MCSS and the top one shows that at 4000 MCSS. Here, $f_0=0.01$ and $\lambda=15.0$.

\newpage
\setlength{\unitlength}{0.240900pt}
\ifx\plotpoint\undefined\newsavebox{\plotpoint}\fi
\sbox{\plotpoint}{\rule[-0.200pt]{0.400pt}{0.400pt}}%
%

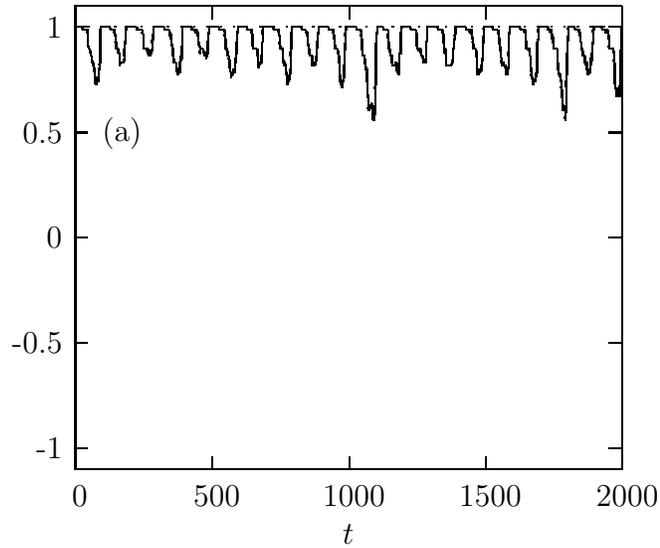
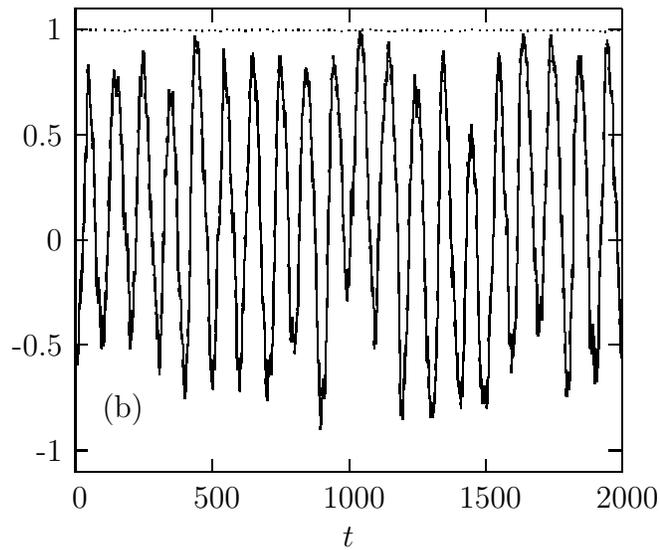
\noindent {\bf Fig-4.} $m_b(t)$ (solid line) and $m_s(t)$
(dotted line) are plotted 
against time $t$. (a) $h_0=2.5$, $\lambda=15.0$ and $T=0.5$ and (b) $h_0=2.5$,
$\lambda=15.0$ and $T=1.25$. 

\newpage
\setlength{\unitlength}{0.240900pt}
\ifx\plotpoint\undefined\newsavebox{\plotpoint}\fi
\sbox{\plotpoint}{\rule[-0.200pt]{0.400pt}{0.400pt}}%
%

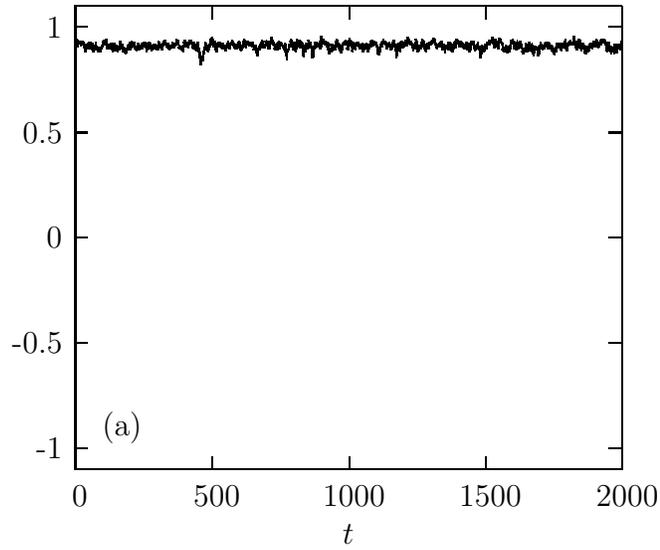
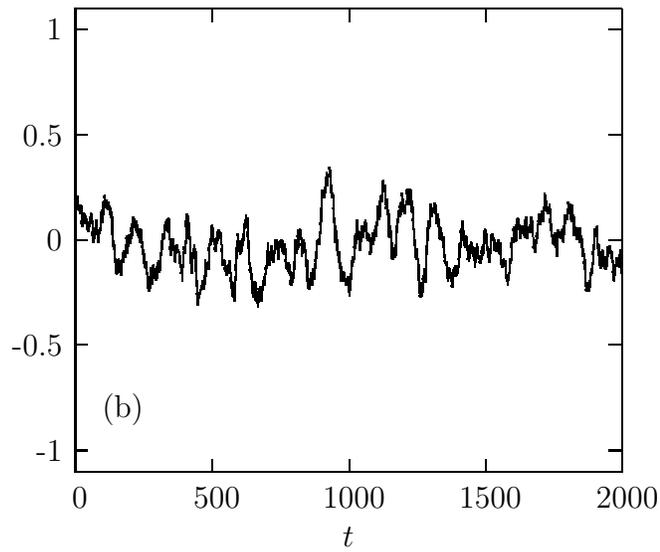
\noindent  {\bf Fig-5.} $m_s(t)$
is plotted 
against time $t$. (a) $h_0=2.5$, $\lambda=15.0$ and $T=2.0$ and (b) $h_0=2.5$,
$\lambda=15.0$ and $T=2.5$. 
\newpage
\setlength{\unitlength}{0.240900pt}
\ifx\plotpoint\undefined\newsavebox{\plotpoint}\fi
\sbox{\plotpoint}{\rule[-0.200pt]{0.400pt}{0.400pt}}%
%

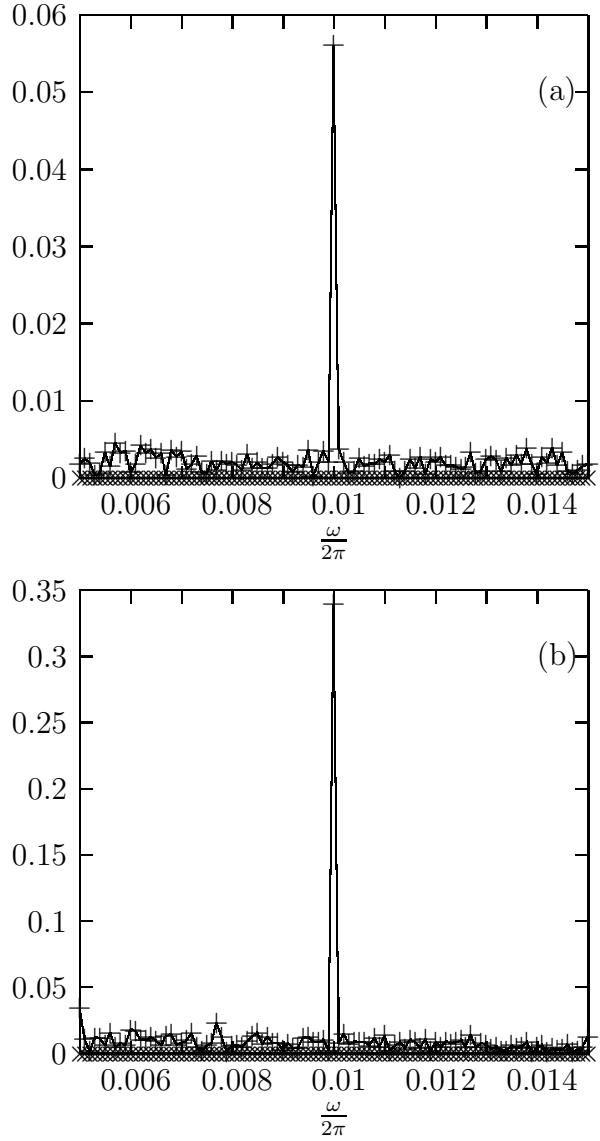
\noindent {\bf Fig-6.} The power spectra for breathing transition.
$|{\hat m_b(\omega)|} (+) $ and $|{\hat m_s(\omega)}| (\times)$ plotted against ${{\omega} \over {2\pi}}$,
for (a) $h_0=2.5$, $\lambda=15.0$ and $T=0.5$ and (b) $h_0=2.5$, $\lambda=15.0$ and $T=1.25$.

\newpage

\setlength{\unitlength}{0.240900pt}
\ifx\plotpoint\undefined\newsavebox{\plotpoint}\fi
\sbox{\plotpoint}{\rule[-0.200pt]{0.400pt}{0.400pt}}%
%

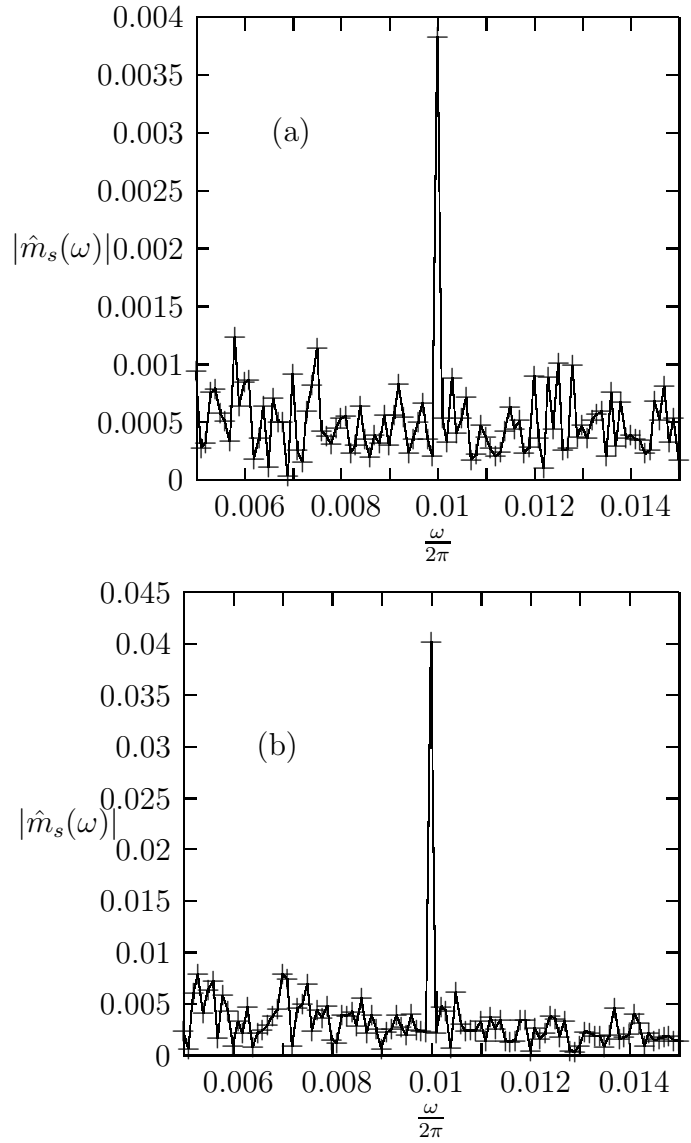
\noindent {\bf Fig-7.} The power spectra for spreading transition.
for (a) $h_0=2.5$, $\lambda=15.0$ and $T=2.0$ and 
(b) $h_0=2.5$, $\lambda=15.0$ and $T=2.5$.

\newpage
\setlength{\unitlength}{0.240900pt}
\ifx\plotpoint\undefined\newsavebox{\plotpoint}\fi
\sbox{\plotpoint}{\rule[-0.200pt]{0.400pt}{0.400pt}}%
%

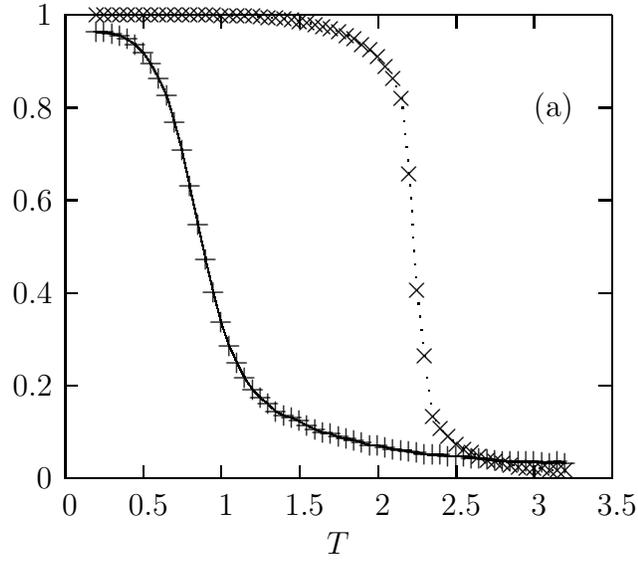
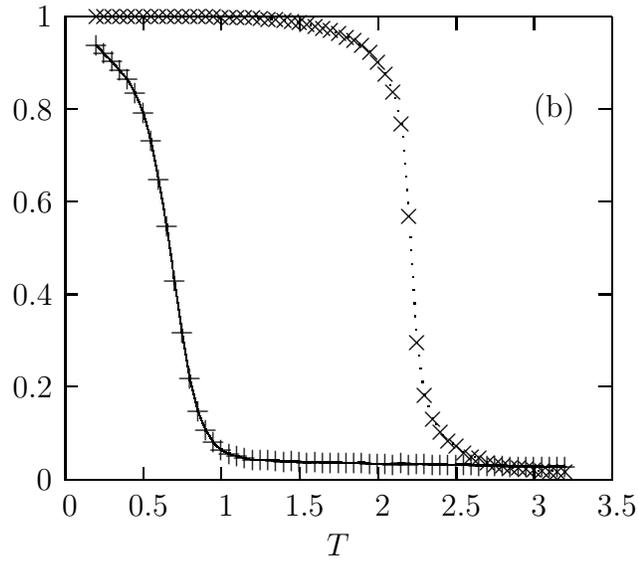
\noindent {\bf Fig-8.} The plots of $Q_b (+)$ and $Q_s(\times)$ 
as a function of temperature
$T$. (a) $h_0=2.5$, $f=0.01$ and $\lambda=15.0$ and (b) $h_0=3.5$, $f=0.01$ 
and $\lambda=15.0$.

\newpage
\setlength{\unitlength}{0.240900pt}
\ifx\plotpoint\undefined\newsavebox{\plotpoint}\fi
\sbox{\plotpoint}{\rule[-0.200pt]{0.400pt}{0.400pt}}%
%

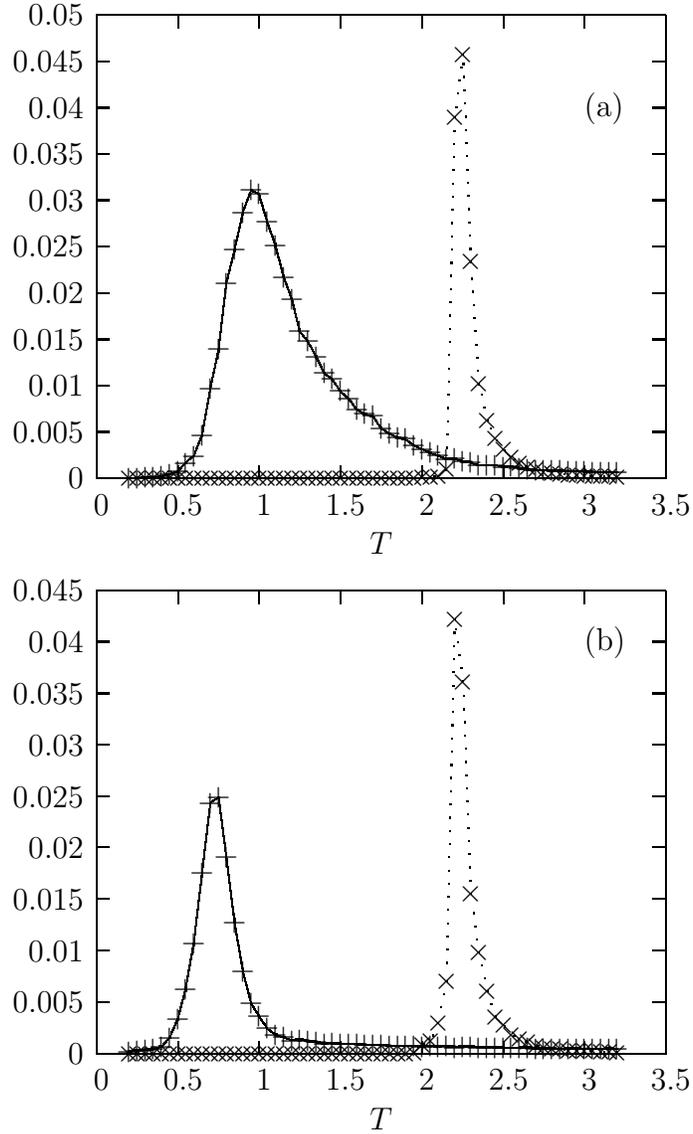
\noindent {\bf Fig-9.} Fluctuations of $Q_b (+)$ and $Q_s (\times)$ plotted
against temperature ($T$) for two different values of field amplitude ($h_0$)
of propagating field. (a) $h_0 = 2.5$ and (b) $h_0 = 3.5$. Here, $\lambda
=15.0$ and $f=0.01$.
\newpage
\setlength{\unitlength}{0.240900pt}
\ifx\plotpoint\undefined\newsavebox{\plotpoint}\fi
\sbox{\plotpoint}{\rule[-0.200pt]{0.400pt}{0.400pt}}%
%

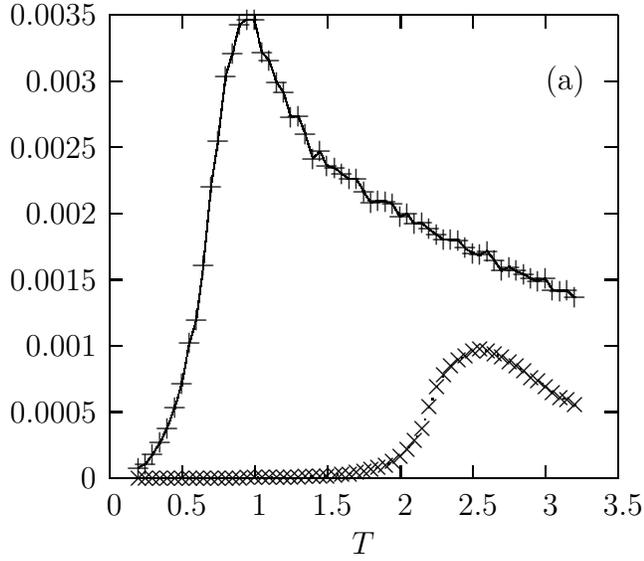
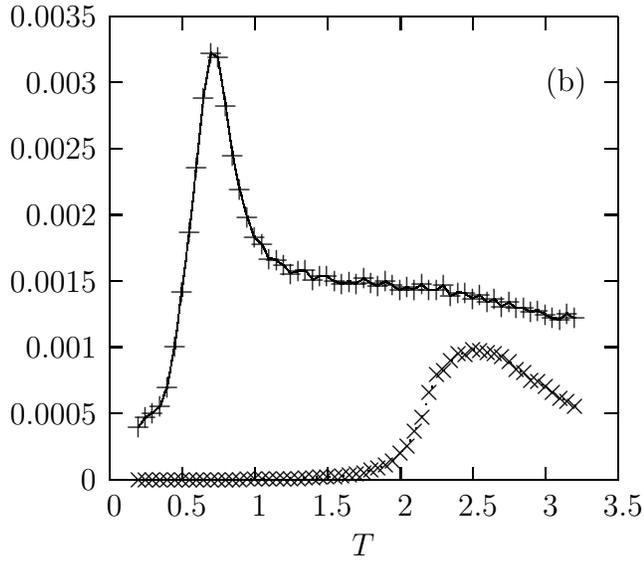
\noindent {\bf Fig-10.} The plots of integrated powers, i.e.,
$|P_b|(+$, for breathing) and 
$|P_s|(\times$, for spreading) as functions of 
temperature ($T$). (a) $h_0=2.5$ and (b) $h_0=3.5$.

\newpage
\setlength{\unitlength}{0.240900pt}
\ifx\plotpoint\undefined\newsavebox{\plotpoint}\fi
\sbox{\plotpoint}{\rule[-0.200pt]{0.400pt}{0.400pt}}%
%

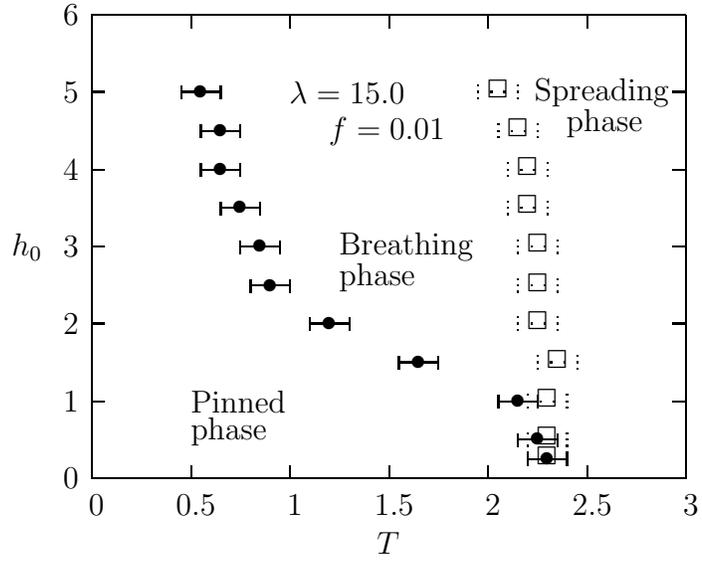
\noindent {\bf Fig-11.} The phase diagram for symmetry breaking dynamic
breathing and spreading transitions. Here, $f_0=0.01$ and $\lambda=15.0$.
\end{document}